\documentstyle[12pt, epsfig]{article}

\def\hb{\hbar}  %
\def\al{\alpha}

\def\rh{\rho}

\def\si{\sigma}

\def\ph{\phi}

\def\om{\omega}
\def\Ga{\Gamma}

\def\Om{\Omega}

\def\fr#1#2{{{#1} \over {#2}}}
\def\prt{\partial}
\def\pr#1{{#1}^\prime} %
\def\ap{\al^\prime}

\def\ket#1{|{#1}\rangle}

\def\abs#1{\left|{#1}\right|}
\def\vect#1{{\bf #1}}

\def\half{{\textstyle{1\over 2}}}

\def\frac#1#2{{\textstyle{{#1}\over {#2}}}}
\def\ni{\noindent}
\def\lsim{\mathrel{\rlap{\lower4pt\hbox{\hskip1pt$\sim$}}
    \raise1pt\hbox{$<$}}}
\def\gsim{\mathrel{\rlap{\lower4pt\hbox{\hskip1pt$\sim$}}
    \raise1pt\hbox{$>$}}}
\def\sqr#1#2{{\vcenter{\vbox{\hrule height.#2pt
         \hbox{\vrule width.#2pt height#1pt \kern#1pt
         \vrule width.#2pt}
         \hrule height.#2pt}}}}

\def\z{{\bf\hat z}}
\def\d#1{{#1}^{\dag}}
\def\ac#1{\left\{{#1}\right\}} %
\def\phihat{{\bf\hat\phi}} %
\def\sect#1{\vglue 0.6cm {\bf\ni {#1}} \vglue 0.4cm} %
\def\Lgr#1#2#3{L_{\scriptstyle #1}^{(\scriptstyle #2)}\left(#3\right)}
\def\Gam#1{\Ga\left({\textstyle #1}\right)} %
\def\alg#1#2{\mbox{#1}(#2)} %

\newcommand{\beq}{\begin{equation}}
\newcommand{\eeq}{\end{equation}}
\newcommand{\bea}{\begin{eqnarray}}
\newcommand{\eea}{\end{eqnarray}}
\newcommand{\rf}[1]{(\ref{#1})}
\newcommand{\eq}[1]{Eq.\ \rf{#1}} %
\renewenvironment{thebibliography}[1]
 { \rm
   \begin{list}{\arabic{enumi}.}
    {\usecounter{enumi} \setlength{\parsep}{0pt}
     \setlength{\itemsep}{3pt} \settowidth{\labelwidth}{#1.}
     \sloppy
    }}{\end{list}}

\begin{document}

\titlepage

\vspace*{0cm}
\begin{center}
{
{\bf Superalgebras for the  Penning Trap \\ }
\vglue 1.0cm
{Neil Russell\\}
\bigskip
{\it Physics Department\\}
\medskip
{\it Northern Michigan University\\}
 \medskip
{\it Marquette, MI 49855, U.S.A.\\}

\vglue 0.8cm
}
\vglue 0.3cm

\end{center}

{\rightskip=3pc\leftskip=3pc\noindent

The hamiltonian describing
a single fermion
in a Penning trap
is shown to be supersymmetric
in certain cases.
The supersymmetries of interest occur
when the ratio of the
cyclotron frequency to the
axial frequency is $3/2$
and the particle has anomalous magnetic moment
$4/3$ or $2/3$.
At these supersymmetric points,
the spectrum shows uniformly spaced
crossed levels.
The associated superalgebras
are $\alg{su}{2|1}$ and
$\alg{su}{1|1}$.
The phase space for this problem
has an
$\alg{osp}{2|6}$
structure
and contains  all
the degeneracy superalgebras.

}

\vfill
\newpage

\baselineskip=20pt

\sect{1. Introduction}
The Penning trap
\cite{penning,dehmelt}
is an impressive
tool for precision spectroscopy
of charged particles.
High-precision measurements
conducted on particles in
a Penning trap
include a comparison of
the anomalous magnetic moments for
the electron and positron
to a precision of $10^{-12}$
\cite{vandyck},
a measurement of
the charge-to-mass ratio
for protons and antiprotons
to $10^{-10}$
\cite{gabrielse95},
and a search for time dependence
in the anomaly frequency of a trapped electron
\cite{rm99}.
Comparable precisions
have been attained in
measurements of
the mass ratio of
the proton to the electron
\cite{farnham},
the masses of molecular ions
\cite{cornell},
and bounds on the anisotropy of space
\cite{prestage}.
Recent theoretical investigations
indicate that
Penning-trap experiments
can constrain Lorentz and CPT violation
at the level of $10^{-20}$
in the context of a general standard-model extension
\cite{bluhm}.
Numerous other applications
of Penning traps exist
\cite{expts}.

In the present paper,
we investigate the symmetries
of the hamiltonian describing
a single charged fermionic particle
confined in a Penning trap
with hyperbolic electrodes.
The symmetry depends on
the relative values of the magnetic and
electric fields
and on the gyromagnetic ratio of the trapped particle.
For certain values of these parameters,
superalgebras
\cite{superalgebras}
arise.

There are relatively few physical manifestations
of superalgebras.
One arises in nuclear physics
\cite{balantekin}.
Another  exists in atomic systems
\cite{kostelecky84-85,kostelecky85},
where a broken quantum-mechanical supersymmetry
has been shown to
underly the properties of
the chemical elements.
It has recently been suggested that
a supersymmetry
also exists in the context of traps
\cite{kostelecky97}.
In this case, a radial supersymmetry
for the trap wave functions
provides a description of a small cloud of particles
in a trap via an effective single-particle formalism.
The associated parallels
between traps and atoms
in the context of quantum-mechanical supersymmetry
have been studied in some detail
\cite{kostelecky85-96}.
Some other results in
quantum-mechanical supersymmetry
are reviewed in
\cite{cooper}.

The supersymmetries  discussed in this paper
for the Penning trap
are of a different type.
The idea is
to consider the full hamiltonian
written in terms of creation and annihilation operators.
The (anti)commutation relations
satisfied by quadratic combinations of
these operators define
the superalgebras relevant to the problem.

In section 2, the relevant features of the Penning
trap are reviewed and some definitions are given.
The relative strengths
of the trapping fields
required for degeneracies to occur
are discussed in section 3.
The central algebra common to all cases
is given in section 4,
and each of the five relevant superalgebras
are presented in turn in sections 5 to 9.
Section 10 summarizes and discusses the results.

%-----------------------------------------
\sect{2. The Penning Trap}
In most situations,
the dynamics of a particle
in a Penning trap
is dominated by its interaction
with a uniform magnetic field
${\bf B}$.
For convenience,
we work in cylindrical coordinates
$(\rh, \ph, z)$
with
$\vect B = B \z$.
A suitable choice of vector potential is
$\vect A = (B \rho/2)\phihat$.

The quadrupole electric field
of the trap
is produced by electrodes
in one of several possible forms
\cite{brown,electrodes}.
We restrict attention to
the case with electrode surfaces
given
in cylindrical coordinates
by the expressions
\beq
z^2 = \rh^2 / 2 \pm d^2
\label{pl3}
\quad ,
\eeq
where $d$ is a constant.
The upper equation
is a hyperboloid of two sheets
and describes the endcap surfaces,
which intersect the $z$ axis at $z=\pm d$
and have potential $V/2$.
The remaining electrode surface
has potential $-V/2$
and has shape determined by
the lower sign in
\eq{pl3}.
It is a hyperboloid of one sheet
encircling the $z$ axis
with waist radius $\sqrt 2 d$
in the $z=0$ plane.
The electrostatic potential
is
\beq
\phi(\rh,\ph,z) = \fr{V}{2d^2}(z^2- \rho^2/2)
\label{pl2}
\eeq
in the trapping region.

Let the trapped particle have
charge $q$
and mass $m$.
We assume that
$q$ and $V$ have the same sign,
thereby ensuring  axial trapping.
Defining the axial frequency
$\omega_z = (q V/ m d^2)^{1/2}$
and the cyclotron frequency
$\omega_c = \abs{q B}/m$,
the hamiltonian
for $q > 0$ is
\beq
\tilde H = -\fr{\hb^2}{2m} \vect{\nabla}^2
  + \frac{1}{8} m \Omega^2\rho^2
  + \half m \omega_z^2 z^2
  + \half \hbar \omega_c i \fr{\prt}{\prt\phi}
\label{pt3}
\quad ,
\eeq
where
$\Omega = (\omega_c^2-2\omega_z^2)^{1/2}$.
For $q<0$, the last term would be negative.
The algebraic structure of the problem
turns out to be
independent of
the sign of $q$,
and to avoid carrying two signs in the expressions
that follow,
we restrict attention to the case $q>0$.

Equation \rf{pt3} separates
by defining
$\Psi(\rho,\phi,z) \equiv (r_0/\rho)^{1/2}W(\rho)\Theta(\theta,z)$,
where
$r_0=(\hbar/m\omega_c)^{1/2}$.
The equation in $\rho$ is
\beq
\left\{-\fr{\hbar^2}{2m} \fr{d^2}{d\rho^2}
     + \fr{\hbar^2}{2m}\fr{(\hat{M}^2-\frac 1 4)}{\rho^2}
     + \frac{1}{8} m \Omega^2\rho^2
     - \left[E-(\hat{K}+\half)\hbar\omega_z
        +\half \hat{M}\hbar\omega_c \right]
\right\} W(\rho)=0
\quad ,
\label{pt4}
\eeq
where $\hat{M}$ and $\hat{K}$ are separation constants
taking values $\hat{M}=0, \pm 1, \pm 2, \ldots$
and $\hat{K}=0,1,2,\ldots$.
The energy eigenvalues $E$ for this problem are
\beq
E_{N,\hat{K},\hat{M}} = \frac{\hbar}{2} \left[\Omega N
            + 2\omega_z \hat{K} - \omega_c \hat{M}
            + (\Omega +\omega_z) \right]
\quad ,
\label{pt5}
\eeq
where $N$ takes values
$N= \hat{\abs M}, \hat{\abs M} +2, \hat{\abs M} +4, \ldots $.
The full solution to the stationary problem
$\tilde{H}\Psi = E \Psi$
involves generalized Laguerre
and Hermite polynomials,
\bea
\Psi_{N,\hat{K},\hat{M}}(\rho,\phi,z) &=& C_{N,\hat{K},\hat{\abs M}}
    \left(\frac{\rho}{r_0}\right)^{\hat{\abs M}}
    \exp{\left[-\frac{k}{4} \left(\frac{\rho}{r_0}\right)^2
        -\half\left(\frac{z}{s_0}\right)^2+i\hat{M}\phi\right]}
\nonumber \\
 && \phantom{somespace} \times\Lgr{N/2-\hat{\abs M}/2}{\hat{\abs M}}
         {\frac{k}{2} \left(\frac{\rho}{r_0}\right)^2}
    H_{\hat K}\left(\frac{z}{s_0}\right)
\quad ,
\label{pt6}
\eea
where $k=\Omega/\omega_c$,
$s_0=(\hbar/m\omega_z)^{1/2}$,
and the normalization coefficient is
\beq
C_{N,\hat{K},\hat{\abs M}}
 = \left[\fr{\sqrt{k}}{r_0^2 \, s_0 \, 2^{\hat K}\pi^{3/2}}
   \left(\fr k 2\right)^{\abs{\hat M} +1/2}
   \fr{\Gam{\frac N 2 - \frac{\hat{\abs M}}{2} + 1}}
         {\Gam{\frac N 2 + \frac{\hat{\abs M}}{2} + 1}
          \Gam{\hat{K}+1}}\right]^\half
\quad . \label{pt7}
\eeq

For the special case
$k=0$,
the coefficient of
the $\rh^2$ term in \eq{pt4} would vanish
and the above solutions
would change.
We exclude this case because
it does not allow long-term confinement.
In the initial stages of trapping
before significant cooling has occurred,
the  motion of the particle can be understood classically.
The possible trajectories are either
circles about the central axis
or curves that exit the trap.
The former are unstable to radial
perturbations.
We therefore restrict attention to the
range of values
$0 < k \leq 1$,
or, equivalently,
$0 < \Om \leq \om_c$.

The hamiltonian
$\tilde H$
can be expressed
in terms of creation and annihilation operators.
A  transformation of the phase space
yields six dimensionless operators
\bea
a, \d a &=& \frac{r_0}{\sqrt{2 k}}
     \left(\pm\prt_x +i\prt_y\right)
    + \sqrt{\frac k 8} \frac 1 {r_0}
     \left(x \pm i y\right) \quad , \nonumber \\
b, \d b &=& \frac{r_0}{\sqrt{2 k}}
     \left(\mp\prt_x+i\prt_y\right)
    - \sqrt{\frac k 8} \frac 1 {r_0}
     \left(x \mp i y\right) \quad , \nonumber \\
c, \d c &=& \pm\frac{s_0}{\sqrt 2} \prt_z
    + \frac 1 {\sqrt 2 \, s_0} z
\label{pt11}
\quad .
\eea
They commute with each other
except for the cases
\beq
[a, \d a]=1, \ \ [b, \d b]=1, \ \ [c, \d c]=1
\quad .
\label{pt13}
\eeq
The transformation \rf{pt11}
preserves the canonical properties
of the phase space,
including the commutation relations
for the momentum and position operators.
Therefore, it is symplectic
\cite{abraham}.

The symplectic transformation
casts the hamiltonian into the form
\beq
\tilde H = \hbar \om_+ (\d a a + \half)
  - \hbar \om_- (\d b b + \half)
  + \hbar \om_z (\d c c + \half)
\label{pt12a}
\quad ,
\eeq
where $\omega_+ = (\omega_c + \Omega)/2$
and $\omega_- = (\omega_c - \Omega)/2$
are called the modified cyclotron frequency
and the magnetron frequency,
 respectively.
The negative sign
in \eq{pt12a}
reveals an inverted oscillator
in the system,
which in principle
could lead to an instability
in the presence of radiation.
However,
in practical situations
this energy loss
is controlled by ensuring
$\om_+ \gg \om_-$,
so particles may be trapped
``indefinitely''
\cite{brown}.

For particles with spin $1/2$,
a term
$\pr H$
must be added to
the hamiltonian
\rf{pt3},
\beq
\pr H \equiv
-\vec{\mu}\cdot\vect{B}
= - \fr g 4 \hbar \omega_c \si_3
\label{pt8}
\quad ,
\eeq
where $g$ is the
Land\'e factor
relating the spin to
the magnetic dipole moment
and $\si_3$ is
the third Pauli matrix.
The operators
$f \equiv (\si_1 + i \si_2)/2$
and
$\d f \equiv (\si_1 - i \si_2)/2$
have one nonzero anticommutation relation,
\beq
\{f, \d f \} \equiv f \d f + \d f f = 1
\label{pt10}
\quad ,
\eeq
and they provide a formalism for describing
the spin degree of freedom.
The additional term in the hamiltonian
is
$\pr H = \hbar \om_g (\d f f - \half),$
where $\om_g= \abs g \omega_c / 2$.
The sign of this term assumes $gq>0$.

Combining the bosonic and fermionic
degrees of freedom
we obtain
the full hamiltonian
$H \equiv \tilde H + \pr H$
in operator form:
\beq
H = \hbar \om_+ (\d a a + \half)
  - \hbar \om_- (\d b b + \half)
  + \hbar \om_z (\d c c + \half)
  + \hbar \om_g (\d f f - \half)
\label{pt12}
\quad .
\eeq

The basis states
for this problem
can be denoted by
$\ket{N_a,N_b,N_c, N_f}$,
where
$N_a, N_b,N_c \in \{0,1,2,\ldots\}$
are the eigenvalues of
the number operators
$\d a a$,  $\ \d b b$ and $\d c c$,
and where
$N_f \in \{0,1\}$
is the eigenvalue
of $\d f f$.

The energy eigenvalues of the system
follow from
\eq{pt12}:
\beq
E(N_a, N_b, N_c, N_f; \om_c, \om_z, g) /\hb
  \equiv  \om_+ (N_a + \half)
  -  \om_- (N_b + \half)
  +  \om_z (N_c + \half)
  +  \om_g (N_f - \half)
\label{pt14}
\quad .
\eeq
The quantum numbers used here are related to the ones
in
\eq{pt5}
by $N_a = (N - \hat{M})/2$,
$N_b = (N + \hat{M})/2$,
and
$N_c=\hat{K}$.

The relative values of the frequencies
in equation
\rf{pt14}
play an important part in
the superalgebra structures
considered below.
To this end,
it is  useful to define
the ratio
$\si$
of the cyclotron and axial frequencies,
\beq
\si \equiv \om_c / \om_z
 = \left( \fr{q B^2 d^2}{m V}\right)^{1/2}
\label{sigma}
\quad .
\eeq
This parameter contains information
about the relative values
of $B$ and $V$.
For experiments
with single trapped electrons,
typical values
\cite{brown}
are
$d \simeq 0.3$ cm,
$B \simeq 6$ T,
and
$V \simeq 10$ V,
giving
$\si \simeq 3\times10^3$.
In this limit of
$\si\gg 1$,
the motion of the trapped particle
is dominated by its interaction
with the magnetic field,
and
\eq{pt14} becomes
\beq
\lim_{\si \rightarrow \infty}
E(N_a, N_b, N_c, N_f; \om_c, \om_z, g) =
\hb\om_c \left[(N_a + \half g N_f)
      -\half \left(\frac{g-2}{2}\right)\right]
\label{bigsigma}
\quad .
\eeq

For experiments with
single trapped protons,
typical values
\cite{brown}
are
$d \simeq 0.1$ cm,
$B \simeq 5$ T,
and
$V \simeq 50$ V,
giving
the lower value $\si \simeq 8$.
As $\si$ is decreased,
the confining effect of
the magnetic field
is weakened,
and trapping becomes impractical
when $\si=\sqrt{2}$.
This corresponds to the excluded case $k=0$.
Exceptional measurement precisions are possible:
for trapped protons,
cyclotron-frequency precisions are at the 90 parts per trillion
level \cite{gabrielse95},
making it feasible to probe minuscule effects
such as Lorentz violation
\cite{violation}.

%-------------------------------------------------------------------
\sect{3. Degeneracy superalgebras and frequency equalities}
The algebraic structures that arise for
the single-particle Penning trap
are superalgebras because
both fermionic and bosonic operators are involved.
We focus on {\em degeneracy} superalgebras
formed from operators
that commute with the hamiltonian,
thereby linking degenerate eigenstates.

All the symmetries we consider
are based on
the hamiltonian
\rf{pt12}.
Superalgebras arise for special
values of the
two parameters
$g$ and $\si$,
which in turn determine the
four characteristic frequencies
$\om_{\pm}$, $\om_z$, and $\om_g$
up to an overall factor.
As an illustrative example,
consider the case of
$g=2/3$ and
$\si = 3/2$.
The Penning-trap hamiltonian is
\beq
H/\hb\om_z= (\d a a + \d c c+1)- \frac 1 2 (\d b b - \d f f + 1)
\quad ,
\label{nx1}
\eeq
and there are two distinct frequencies,
$\om_+ = \om_z = 2 \om_- = 2 \om_g$.
The generator $\d b f$
increases $N_b$ by one unit
while decreasing $N_f$ by the same amount.
It commutes with the hamiltonian
because of the equality of $\om_g$ and $\om_-$.

In the most general case,
$\om_{\pm}$, $\om_z$, and $\om_g$
are distinct.
There are four generators
constructed from quadratic combinations
of creation and annihilation operators
that commute with the hamiltonian:
$\d a a$, $\d b b$, $\d c c$,
and $\d f f$.
They generate an abelian algebra
$\alg u 1 \times \alg u 1 \times \alg u 1 \times \alg u 1 $
and form a complete set
of commuting operators.
Their interpretation as constants of the motion
is  considered in the next section.
The generators of this abelian algebra
commute with the hamiltonian
and with any other degeneracy operators
regardless of the values of $g$ and $\si$.
Therefore, all the degeneracy superalgebras
considered below
contain this four-dimensional central algebra.

Even with four distinct frequencies,
degeneracies in the energies can occur.
Consider the case of $\si=9/4$ and $g=2/3$.
The corresponding hamiltonian is
\beq
H / \hb\om_z = 2(\d a a + \half)-\frac 1 4 (\d b b + \half) +
       (\d c c + \half) + \frac 3 4 (\d f f - \half)
\label{ratfreq}
\quad .
\eeq
The point is that the associated
frequencies are all rational multiples of each other.
By taking combinations higher than quadratic
in the creation or annihilation operators,
generators can be constructed that
commute  with the hamiltonian.
Take, for example, the operator $\d a c^2$.
It increases $N_a$ by one unit
and decreases $N_c$ by two units.
This ensures
commutation with the hamiltonian
because the associated frequencies $\om_+$ and $\om_z$
are in the ratio $2:1$.
Other generators that commute
with this hamiltonian are
$a (\d c)^2$, $(\d b)^4 \d c$, $b^4 c$, $a b^8$,
$\d a (\d b)^8$, and $bc \d f$.
A detailed study of
the algebraic structures associated with
cubic and higher combinations
of creation or annihilation operators
lies beyond the scope of the present work.

Next,
consider the case of three distinct frequencies.
For a superalgebra to arise,
$\om_g$ must be equated with
another frequency.
We give a few examples.
For
$\si=9/4$ and $g=2/9$, we find that the ratio
$\om_+ : \om_- : \om_z : \om_g$ is $8:1:4:1$,
so that $\om_g=\om_-$.
For
$\si=11/6$ and $g=18/11$, the frequency ratio is
$9:2:6:9$,
so that $\om_g=\om_+$.
For
$\si=9/4$ and $g=8/9$, the frequency ratio is
$8:1:4:4$,
so that $\om_g=\om_z$.
The superalgebras that arise are all isomorphic and
are discussed in section 5.

Next, consider
ways in which
the single-particle Penning-trap system
can have two distinct
characteristic frequencies
in a rational ratio.
Of these, we focus on the simplest possible ratio,
$2:1$.
It turns out that there are only two cases.
One arises for
$g=2/3$ and $\si=3/2$
and the corresponding hamiltonian is given in
\eq{nx1}.
This case is considered in section 6.
The other arises
for $\si=3/2$ and $g=4/3$.
It is
the intersection point of the curves
$\om_+$, $\om_z$, and $\om_g$
as functions of $\si$,
and is illustrated in
Figure 1.
For this case,
the frequencies are
$\om_+ = \om_z = \om_g = 2 \om_-$
and the associated supersymmetries
are considered in detail in section 7.

It is not possible to equate all four
frequencies
to yield a single  characteristic
frequency for the system.
This can be seen in
Figure 1,
which shows that
$\om_z$ cannot equal $\om_-$.

The two cases with two distinct characteristic
frequencies are special.
They represent the largest possible
superalgebras that
can be constructed
from quadratic generators for
the single-particle Penning trap.
Both cases have $\si=3/2$,
but differ in the values of $g$.

%----------------------------------------------------------

\sect{4. Constants of the motion for the supersymmetric configuration}
For the supersymmetric point $\si=3/2$,
the hamiltonian can be written
in terms of four constants of the motion
$H_\rho$, $H_\phi$, $H_z$, and $H_f$
to be defined below:
\beq
H = H_\rho + H_\phi + H_z + H_f
\quad .
\label{conH}
\eeq
These operators
have simple physical interpretations.

The first one is the energy operator
of a harmonic oscillator
in the $xy$ plane
with frequency
$\om_z/4$:
\bea
H_\rho &\equiv& - \fr{\hb^2}{2m}\left(\prt_{\rho}^2
         + \fr 1 {\rh} \prt_{\rh}
         + \fr 1 {\rho^2}\prt_{\phi}^2\right)
   + \fr 1 2 m \left(\fr{\om_z}{4}\right)^2 \rho^2
\nonumber \\
   &=& \fr {\hb\om_z}{4}\left(\d a a + \d b b + 1\right)
\quad .
\label{con1}
\eea

The operator $H_\phi$ is a rotational energy about the $z$ axis:
\beq
H_\phi \equiv \fr 1 2 \hb \om_c i \prt_\phi
    = - \fr 1 2 \om_c L_z
\quad ,
\label{con2}
\eeq
where $\om_c=3\om_z/2$.
This term has negative eigenvalues
for $L_z$ in the $+z$ direction.
This is consistent with the presence of
an inverted harmonic oscillator in the Penning trap.
The angular momentum about the $z$ axis
can be expressed in terms of the
creation and annihilation
operators
\cite{biedenharn}
as
\beq
L_z = \hb ( \d b b - \d a a )
\quad .
\label{conL}
\eeq

The operator $H_z$ is the energy operator
of a harmonic oscillator with frequency
$\om_z$ on the $z$ axis:
\beq
H_z \equiv - \fr{\hb^2}{2m} \prt_z^2 + \fr 1 2 m \om_z^2 z^2
  = \hb \om_z \left(\d c c + \frac 1 2 \right)
\quad .
\label{con3}
\eeq

The operator $H_f$ is the energy operator
for the splitting between the two spin projections
onto the $z$ axis:
\beq
H_f \equiv \hb \om_g \left(\d f f -\fr 1 2 \right)
\quad .
\label{con4}
\eeq

The four operators
$H_\rho$, $H_\phi$, $H_z$ and $H_f$
form an alternative
complete set of commuting operators
for the single-particle Penning trap.
They form a basis of the abelian center of all
the degeneracy superalgebras for this system,
and their associated energies are independent of each other.

%----------------------------------------------------

\sect{5. Three distinct frequencies}
For this case,
$\om_g$ must equal one of the other frequencies
and the remaining two frequencies
must each be distinct from this value
and from each other.
This can occur in numerous ways.
As an example,
consider the case with
$\si = 11/6$ and $g=18/11$ mentioned in section 3.
The hamiltonian is
\beq
H/\hb\om_z = \frac 3 2 (\d a a + \d f f)
           - \frac 1 3 (\d b b +\frac 1 2)
           + (\d c c + \frac 1 2)
\quad .
\label{thr1}
\eeq
Define the operators
\beq
\begin{array}{lcl}
J &\equiv& \d a a + \d f f
\quad , \nonumber\\
\overline J &\equiv& \d a a - \d f f + 1
\quad , \nonumber\\
F_{+1} &\equiv& \d a f
\quad , \nonumber\\
F_{-1} &\equiv& a \d f
\quad . \nonumber
\end{array}
\label{thr2}
\eeq
Note from
\eq{pt11}
that they depend on
the value of $k$,
and that for this case
$k =\sqrt{\si^2-2}/\si = 7/11$.
They commute with the hamiltonian
and generate a superalgebra.
The only nonzero relations are
\beq
[\overline J, F_{\pm 1}] = \pm 2 F_{\pm 1}
\quad ,
\quad
\{F_{+1}, F_{-1}\} = J
\quad .
\label{thr3}
\eeq
This algebra has a nontrivial ideal spanned by
$J, F_{\pm 1}$
and so is not simple.
The ideal is the nilpotent superalgebra $\alg{su}{1|1}$
with Lie part $\alg{u}{1}$
generated by $J$.
The operator $\overline J$ does not commute with the odd operators
$F_{\pm 1}$,
so we denote the superalgebra by
$\alg{u}{1}\oslash\alg{su}{1|1}$
to indicate the absence of a direct product.

The full degeneracy algebra for the hamiltonian
\eq{thr1} includes elements which complete the basis of the center.
The structure is
$\alg u 1 \times \alg u 1 \times \alg u 1 \oslash \alg{su}{1|1}$,
generated by $\d b b$, $\d c c$, $\overline J$,
and $\{J, F_{\pm 1}\}$.

Given a pair $F_{\pm 1}$
of mutually hermitian-conjugate generators,
(self-)hermitian generators are obtained
by the combinations $T_1 = (F_{+1} + F_{-1})/2$
and $T_2 = i(F_{+1} - F_{-1})/2$.
We define
nonhermitian ladder generators
because they are useful for calculations.
The actual hermitian generators
within the superalgebras
can always be constructed by this method.

Another way to obtain a supersymmetry
with three distinct frequencies in the system
is to set $\om_g = \om_-$.
Consider the example mentioned in section 3
with $g= 2/9$ and $\si= 9/4$,
which corresponds to $k=7/9$.
The hamiltonian is
\beq
H/\hb\om_z = 2 (\d a a + \frac 1 2)
           - \frac 1 4 (\d b b - \d f f + 1)
           + (\d c c + \frac 1 2)
\quad .
\label{thr4}
\eeq
We define four operators that
commute with the hamiltonian:
\beq
\begin{array}{lcl}
K &\equiv& \d b b + \d f f
\quad , \nonumber\\
\overline{K} &\equiv& \d b b - \d f f + 1
\quad , \nonumber\\
F_{+2} &\equiv& \d b \d f
\quad , \nonumber\\
F_{-2} &\equiv& b f
\quad . \nonumber
\end{array}
\label{thr5}
\eeq
They generate a superalgebra with
nonzero relations
\beq
[K, F_{\pm 2}] = \pm 2 F_{\pm 2}
\quad ,
\quad
\{F_{+2}, F_{-2}\} = \overline{K}
\quad .
\label{thr6}
\eeq
Comparison of these relations with those in
\eq{thr3} shows that the two algebras are isomorphic.
The full superalgebra for this example is
$\alg u 1 \times \alg u 1 \times \alg u 1 \oslash \alg{su}{1|1}$,
generated by $\d a a$, $\d c c$, $K$,
and $\{\overline{K}, F_{\pm 2}\}$.

One might expect different superalgebras
to arise for the hamiltonians
\rf{thr1} and
\rf{thr4}
because of the opposite signs of
$\d a a$ and $\d b b$ relative to $\d f f$.
However, this is not the case, and
the isomorphism relating the operators in \eq{thr2} and \eq{thr5}
is given explicitly by
\beq
a \leftrightarrow b \quad , \quad
\d a \leftrightarrow \d b \quad , \quad
f \leftrightarrow \d f \quad .
\label{thr7}
\eeq
It follows from this observation that the {\em only}
superalgebra that can arise for three distinct frequencies is
$\alg u 1 \oslash \alg{su}{1|1}$.
In all cases of this type,
the full symmetry is
$\alg u 1 \times \alg u 1 \times \alg u 1 \oslash \alg{su}{1|1}$.

This supersymmetry is relevant
to experiments
with electrons or positrons,
where $\si\gg1$ and $g\simeq 2$.
Taking $g=2$,
the hamiltonian for $\si \gg 1$ is
\beq
H/\hb \approx \om_c(\d a a + \d f f) + \om_z (\d c c + \frac 1 2)
               - \om _-(\d b b + \frac 1 2)
\quad ,
\label{thr8}
\eeq
with $\om_c \gg \om_z \gg \om_-$.
However,
the supersymmetry is broken
because in the physical situation
$g$ is  slightly larger than two,
so $\om_+$ is always slightly less than $\om_g$
no matter how strong the magnetic field.
The value of the $g$ factor determines
the degree to which this supersymmetry is broken
in the strong-$B$
limit.
In this regime,
the particle experiences
a uniform magnetic field
and has associated supercoherent
states
\cite{fkt91}.
If $g$ were exactly equal to $2$,
the anomaly
$a_e = (g-2)/2 \simeq 10^{-3}$
would be zero,
and the spin-up and spin-down ladders
would have no relative energy
shift
\cite{ew80}.

%-------------------------------------------
\sect{6. Two pairs of equal frequencies:
$\om_+ = \om_z = 2\om_- = 2\om_g$}
For
$g=2/3$ and
$\si = 3/2$,
the Penning-trap hamiltonian
is given in \eq{nx1}.
Four linearly independent generators
constructed from $a, \d a, c$ and $\d c$
that commute with this hamiltonian
are
\beq
\begin{array}{c}
\overline{L} \equiv \d a a + \d c c + 1 \; ,  \\
L \equiv \half (\d a a - \d c c) \; ,  \
E_{+2} \equiv \d a c \; ,  \  E_{-2} \equiv a \d c  \; .
\end{array}
\label{nx2}
\eeq
The generator $\overline{L}$ commutes with
the other three,
forming a $\alg{u}{1}$ subalgebra.
The generators $E_{+2}$ and $E_{-2}$
are hermitian conjugates
and are themselves non-hermitian.
They are ladder operators,
which  together with $L$
give the Lie algebra $\alg{so}{3}$:
\beq
[L, E_{\pm 2}]  =  \pm  E_{\pm 2} \; ,  \quad
[E_{+2}, E_{-2}] =   2 L \; .
\label{nx3}
\eeq
The remaining generators
of the superalgebra are
$K$, $\overline{K}$, and $F_{\pm 2}$
defined in \eq{thr5},
but with $k=1/3$.
They span the superalgebra
$\alg u 1 \oslash \alg{su}{1|1}$
with nonzero relations given in
\eq{thr6}.

Combining the relations of
\eq{nx3} and \eq{thr6},
the full degeneracy superalgebra
for the hamiltonian in \rf{nx1}
is $\alg u 1 \times \alg{so}{3}
\times \alg u 1 \oslash \alg{su}{1|1}$,
generated by
$\overline{L}$,
$\{L,E_{\pm 2}\}$,
$\overline{K}$,
and
$\{K, F_{\pm 2}\}$.
It is implicit that
for this case $k=7/9$
in the definitions
\rf{thr5}.

For $gq<0$,
the second  term of the hamiltonian
\rf{nx1}
becomes $-(\d b b + \d f f )/2$.
A full set of generators that
commute with the hamiltonian
is obtained from
\eq{nx2}
and by making the replacements
$f \rightarrow \d f$ and
$\d f \rightarrow f$
in \eq{thr5}.
This operation is an automorphism,
leaving  the relations
\rf{thr6}
and
\rf{nx3}
unchanged.

From a given state
$\ket{N_a,N_b,N_c,N_f}$,
the elements defined
in  Eqs. \rf{thr5} and \rf{nx2}
generate all the states
in the  degenerate subspace.
The Lie algebra $\alg{so}{3}$
generates states differing in
the $N_a$ and $N_c$ eigenvalues.
For example,
\beq
E_{+2} \ket{N_a,N_b,N_c,N_f}
\sim \ket{N_a+1,N_b,N_c-1,N_f}
\quad .
\label{nx7a}
\eeq
In contrast,
the subsuperalgebra $\alg{su}{1|1}$
acts to give states differing only in
$N_b$ and $N_f$.
For example,
\beq
F_{+2} \ket{N_a,N_b,N_c,N_f}
\sim ((N_f+1) \bmod 2) \ket{N_a, N_b+1, N_c, (N_f+1) \bmod 2}
\quad .
\label{nx7b}
\eeq

Insight into the physical implications
of the superalgebra can be gained from Figure 2.
It  plots the energy levels
of the Penning trap
versus $\si$
for the states with quantum numbers
$N_a=0,1,2$,
$N_b=0 \ldots 3$,
$N_c=0,1$,
and
$N_f=0,1$.
At
$\si=3/2$,
the hamiltonian has the form of
\eq{nx1}.
The coefficients of the two terms
show that the frequencies
are in the ratio 2:1.
This gives the uniform spacing
of the energy levels
and creates the sharply defined crossing features
at this supersymmetry point
on the plot.

Figure 2 also reveals the set of
evenly spaced degenerate levels
at $\si=2.25$,
for which the hamiltonian has the form in \eq{ratfreq}.

The operators $L$, $\overline{L}$, $K$, and
$\overline{K}$ form a complete set of commuting operators
for the system.
They can be expressed in terms of
the more physical operators defined in section 4:
\bea
\hb \om_z \overline{L} &=& 2 H_\rho + \frac 2 3 H_\phi + H_z
\quad , \\
\hb \om_z L &=& H_\rho + \frac 1 3 H_\phi - \frac 1 2 H_z
\quad , \\
\hb \om_z \overline{K} &=& 2 H_\rho - \frac 2 3 H_\phi - 2 H_f
\quad , \\
\hb \om_z K &=& 2 H_\rho + \frac 2 3 H_\phi + 2 H_f
\quad .
\eea

\sect{7. Three equal frequencies:
      $\om_+=\om_z=\om_g=2\om_-$}
Three frequencies can  be equated by setting
$g=4/3$ and
$\si= 3/2$,
giving the hamiltonian
\beq
H/\hb\om_z=(\d a a+ \d c c + \d f f + \half)
   -\frac 1 2 (\d b b + \half)
 \equiv M - \frac 1 2  \overline M
\quad .
\label{pta8}
\eeq
The generators $M$ and $\overline M$,
defined by the expressions in parentheses,
commute with each other
and with $H$.
They therefore form an independent
$\alg u 1 \times \alg u 1$
subalgebra of the full degeneracy superalgebra.

In addition to $M$ and $\overline M$,
there are four independent even elements
given by
\beq
\tilde L \equiv  \half (\d a a + \d c c) + \d f f
\label{pta9}
\eeq
and by
$L, \, E_{\pm2}$ defined in \eq{nx2}.
The generator
$\tilde L$ commutes with the even elements
$L, E_{\pm 2}$,
which in turn
satisfy the commutation relations \rf{nx3}
for the compact Lie algebra $\alg{su}{2}$.

There are four odd elements
that commute with the hamiltonian:
$F_{\pm 1}$ as defined in \eq{thr2}
but with $k=1/3$,
and
\beq
F_{+3} \equiv \d c f \; , \quad
F_{-3} \equiv c \d f   \; .
\label{pta10}
\eeq
Their nonzero anticommutation relations are
\beq
\ac{F_{+1},F_{-1}} = \tilde L + L  \; , \quad
\ac{F_{+3},F_{-3}} = \tilde L - L  \; , \quad
\ac{F_{\pm 1},F_{\mp 3}} =  E_{\pm 2} \; ,
\label{pta13}
\eeq
and their nonzero commutation relations
with the even elements are
\beq
\begin{array}{ll}
[\tilde L,F_{\pm 1}] = \mp \half F_{\pm 1} \; , &
[\tilde L,F_{\pm 3}] = \mp \half F_{\pm 3} \; , \\[2mm]
[L,F_{\pm 1}] = \pm \half F_{\pm 1} \; , &
[L,F_{\pm 3}] = \mp \half F_{\pm 3} \; , \\[2mm]
[E_{\pm 2},F_{\pm 3}] = \pm F_{\pm 1} \; ,   &
[E_{\pm 2},F_{\mp 1}] = \mp F_{\mp 3} \; .
\end{array}
\label{pta12}
\eeq

The superalgebra
with generators given in
Eqs. \rf{nx2}, \rf{pta9}, and \rf{pta10} is
$\alg{su}{2|1}$,
with Lie part
$\alg{u}{1} \times \alg{su}{2}$.
The first component is  generated by
$\tilde L$
and the second by
$\{L, E_{\pm 2}\}$.
The action of these generators
on the eigenstates of the hamiltonian
is similar to that displayed in
Eqs. \rf{nx7a} and \rf{nx7b},
except that here
the values of
$N_a$, $N_c$ and $N_f$
are affected.

The full degeneracy structure
of the hamiltonian \rf{pta8} is
$\alg u 1 \times \alg u 1 \times \alg{su}{2|1}$.
It has three subalgebras,
generated by the sets
$\{M\}$, $\{\overline M\}$, and
$\{L, \tilde L, E_{\pm 2}, F_{\pm 1}, F_{\pm 3}\}$.

The hamiltonian for $gq<0$
is found by replacing
$(\d f f-1/2) \rightarrow -(\d f f - 1/2)$
in \rf{pta8}.
To obtain the generators commuting with
this hamiltonian,
the replacements
$f \rightarrow \d f$ and
$\d f \rightarrow f$ are made
in the definitions
for all the operators.
This automorphism leaves unchanged
the superalgebra relations.
Thus,  the algebraic structure
is again independent of the sign of
$gq$
for the trapped particle.

Figure 3 plots the energy levels
versus $\si$ for
the states with quantum numbers
$N_a=0,1,2$,
$N_b=0,1,2$,
$N_c=0,1$,
and
$N_f=0,1$.
Because the frequencies are
in a rational ratio,
the supersymmetry point
has uniformly spaced crossings at
$\si=3/2$.

The operators $\overline M$, $M$, $\tilde L$, and
$L$ form a complete set of commuting operators
for the system.
They can be expressed in terms of the alternative basis
of section 4:
\bea
\hb \om_z \overline M &=& 2 H_\rho - \frac 2 3 H_\phi
\quad , \\
\hb \om_z M &=&  2 H_\rho + \frac 2 3 H_\phi + H_z + H_f
\quad , \\
\hb \om_z \tilde L &=& H_\rho + \frac 1 3  H_\phi +
    \frac 1 2  H_z + H_f \quad , \\
\hb \om_z L &=& H_\rho + \frac 1 3 H_\phi - \frac 1 2 H_z
\quad .
\eea
These expressions can be inverted.
For example,
the spin-splitting operator $H_f$ can be shown to be
$H_f = \hb \om_z (2 \tilde L - M)$.

\sect{8. Hypothetical case of four equal frequencies}

The largest possible degeneracy superalgebra
in a system of the form of \rf{pt12}
would arise if all the frequencies
could be set equal.
No choices of $g$ and $\si$ allow this
in the Penning trap,
as can be seen from Figure 1.
Nonetheless,
it is of interest to consider
the degeneracy superalgebra that would arise
from a hamiltonian of the form
\beq
H_0 \equiv  \d a a
  -  \d b b
  +  \d c c
  +  \d f f
\label{ga1}
\quad ,
\eeq
where $f$ and $\d f$ are fermionic
and the other operators are bosonic,
because this superalgebra
contains all the superalgebras
discussed in sections 5, 6,  and  7
as subsuperalgebras.
This superalgebra is
$\alg u 1 \times \alg{su}{2,1|1}$,
as shown below.

The hamiltonian $H_0$
forms an independent $\alg u 1$ subalgebra
by definition.
There are eight other independent generators
commuting with this hamiltonian
that are constructed only from bosonic operators.
Expressed in the Cartan-Weyl basis, they are
$E_{\pm2}$ already defined in \eq{nx2},
and
\beq
\begin{array}{ll}
H_1 \equiv \d b b+\d c c+1 \; ,  &
H_2 \equiv \d a a+\d b b+1 \; ,  \\
E_{+1} \equiv \d b \d c \; , &
E_{-1} \equiv b c \; , \\
E_{+3} \equiv \d a \d b \; , &
E_{-3} \equiv a b \; ,
\end{array}
\label{ga2}
\eeq
and they satisfy the
nonzero commutation relations
\beq
\begin{array}{lll}
  [H_1,E_{\pm 1}] = \pm 2 E_{\pm 1} \; ,
& [H_1,E_{\pm 2}] = \mp   E_{\pm 2} \; , &
  [H_1,E_{\pm 3}] = \pm   E_{\pm 3} \; ,
\\[1mm]
  [H_2,E_{\pm 1}] = \pm   E_{\pm 1} \; ,
& [H_2,E_{\pm 2}] = \pm   E_{\pm 2} \; , &
  [H_2,E_{\pm 3}] = \pm 2 E_{\pm 3} \; ,
\\[1mm]
  [E_{\pm 2},E_{\mp 3}]=\mp E_{\mp 1} \; ,
& [E_{\pm 3},E_{\mp 1}]=\mp E_{\pm 2} \; , &
  [E_{\pm 1},E_{\pm 2}]= \mp E_{\pm 3} \; ,
\\[1mm]
  [E_{+1},E_{-1}] = - H_1       \; ,
& [E_{+2},E_{-2}] = - H_1 + H_2 \; , &
  [E_{+3},E_{-3}] = - H_2       \; .
\label{ga5}
\end{array}
\eeq
These generators provide a description of
the Lie algebra su$(2,1)$.

Including the two fermionic operators
$f$ and $\d f$
allows the introduction of
seven more generators
that commute with the hamiltonian \rf{ga1},
of which one,
\beq
H_3 \equiv \d a a - \d b b + \d c c + 3\d f f - 1 \; ,
\label{ga6}
\eeq
is even and commutes with the eight other even generators.
The six others are odd generators
defined earlier:
$F_{\pm 1}$,
$F_{\pm 2}$,
and
$F_{\pm 3}$.
They satisfy anticommutation relations,
of which the only nonzero ones are
\bea
  \ac{F_{\pm 2},F_{\pm 3}} &=& E_{\pm 1} \; , \ \
  \ac{F_{\pm 1},F_{\mp 3}} \ = \ E_{\pm 2}   \; , \ \
  \ac{F_{\pm 1},F_{\pm 2}} \ = \ E_{\pm 3}   \; ,
\nonumber \\
\ac{F_{+1},F_{-1}} &=& - \frac 1 3 H_1 + \frac 2 3 H_2 +
        \frac 1 3 H_3 \; , \nonumber \\
\ac{F_{+2},F_{-2}} &=&   \frac 1 3 H_1 + \frac 1 3 H_2 -
        \frac 1 3 H_3 \; , \nonumber \\
\ac{F_{+3},F_{-3}} &=&   \frac 2 3 H_1 - \frac 1 3 H_2 +
\frac 1 3 H_3 \; .
\label{ga9}
\eea
Note that these anticommutators yield elements
within the even part of the superalgebra,
as expected.
Commutation relations
between even and odd generators
produce generators in
the odd part of the superalgebra.
The nonzero cases are
\beq
\begin{array}{lll}
  [H_3,F_{\pm 1}] = \mp 2 F_{\pm 1} \; ,
& [H_3,F_{\pm 2}] = \pm 2 F_{\pm 2} \; ,
& [H_3,F_{\pm 3}] = \mp 2 F_{\pm 3} \; ,
  \\[1mm]
  [H_1,F_{\pm 2}] = \pm  F_{\pm 2} \; ,
& [H_1,F_{\pm 3}] = \pm  F_{\pm 3} \; ,
& \\[1mm]
  [H_2,F_{\pm 1}] = \pm  F_{\pm 1} \; ,
& [H_2,F_{\pm 2}] = \pm  F_{\pm 2} \; ,
& \\[1mm]
  [E_{\pm 1},F_{\mp 2}] = \mp F_{\pm 3} \; ,
& [E_{\pm 1},F_{\mp 3}] = \mp F_{\pm 2} \; ,
& \\[1mm]
  [E_{\pm 3},F_{\mp 1}] = \mp F_{\pm 2} \; ,
& [E_{\pm 3},F_{\mp 2}] = \mp F_{\pm 1} \; ,
\end{array}
\label{ga12}
\eeq
and the last two relations of \rf{pta12}.
The fifteen-dimensional superalgebra $\alg{su}{2,1|1}$
considered here
has Lie subalgebra
$\alg u 1 \times \alg{su}{2,1}$,
with the first component generated by $H_3$.
The $\alg{su}{2,1}$ subalgebra has eight dimensions,
with basis given in \eq{ga2}.

Including the $\alg u 1$ algebra
generated by $H_0$,
the full degeneracy superalgebra
for the hamiltonian \rf{ga1} is
$\alg u 1 \times \alg{su}{2,1|1}$.

\sect{9. Phase-space superalgebra}
The degeneracy superalgebras
considered above
are subsuperalgebras
of a still larger superalgebra
$\cal{A}$,
where the generators are formed from
{\em all} possible
independent quadratic combinations of
creation or annihilation operators.
This algebra is not a degeneracy superalgebra,
although it contains
the degeneracy superalgebras mentioned
in the previous sections.
In the superalgebra
$\cal{A}$,
there are $12$ odd generators
formed by pairing each of the six bosonic operators
$a, \d a, b, \d b, c, \d c,$
with each of the fermionic operators
$f, \d f$.
There are $21$ even generators
formed from pairs of bosonic operators
including, for example,
$\d a \d a, \d a b, b b, b \d c $.
These generate an
$\alg{sp}{6}$ subalgebra.
A further even generator,
$\d f f$,
is formed from the fermionic operators.
Taken together,
the $34$ generators define the superalgebra
$\alg{osp}{2|6}$,
which has even part
$\alg{sp}{6} \times \alg{so}{2}$.

The $\alg{osp}{2|6}$ superalgebra
$\cal{A}$
is not unique
to the Penning-trap system,
since it would arise for any combination of signs
for the number operators
in the hamiltonian \rf{pt12}.
The point is that
$\cal{A}$
exists even before a potential for
the physical problem is defined.
The only requirement
for $\cal{A}$ to be a relevant algebra
is that the
system describe a single fermion
in a phase space with three space
and three momentum dimensions.
Thus, the superalgebra
$\alg{osp}{2|6}$
describes the properties of the phase space
for the problem.

The hamiltonian for
the Penning trap is fixed
by specifying the parameters
$\om_c$, $\om_z$ and $g$.
For each of the cases in
sections 5, 6, and 7,
the degeneracy superalgebra
is a subsuperalgebra of
the phase-space superalgebra.
We therefore find
a hierarchy of nested superalgebras:
${\cal A}=\alg{osp}{2|6} \supset \alg{so}{2,1|1}
\supset \cal{D}$,
where
$\cal{D}$
is any of the degeneracy superalgebras of
sections 5, 6  or 7.

We have considered only structures
arising from {\em quadratic} combinations
of creation or annihilation
operators.
The issue of
the role played by higher-order combinations,
such as those commuting with
\eq{ratfreq},
is related to
Clifford-algebra theory
\cite{clifford-alg}
but lies outside the scope of this paper.

%------------------------------------------

\sect{10. Summary and Discussion}
Several superalgebras are associated with
the single-particle Penning trap.
The various cases depend
on
the gyromagnetic ratio of the trapped particle
and the relative strengths of the
magnetic and electric fields.
This paper  considers the
degeneracy superalgebras
of operators that commute with the
hamiltonian.
The relevant superalgebras
are summarized in Table 1.

In general,
superalgebra descriptions
might be expected for
trap systems having  energy separations
between spin states
equal to the separations between
the bosonic oscillator-like levels.
This guarantees the existence of odd generators
that commute with the hamiltonian.
Traps in which the spin cannot
to be reversed,
such as the TOP or Ioffe-Pritchard traps
\cite{kostelecky97},
are therefore
unlikely to have
superalgebra structures
of the type described here.
Superalgebras of this kind are also unlikely
for traps where
the spin states are independent of
a magnetic field,
as is the case for the Paul trap
\cite{kostelecky97}.
However,
supersymmetries of another type
do appear in these systems
\cite{kostelecky97}.

Some other issues beyond the scope of
this paper are of potential interest.
In particular,
the spectrum-generating superalgebras
would be relevant
to a complete study of the
properties of the Penning trap.
Furthermore,
higher-rank combinations of operators,
such as those mentioned for
the $\si=9/4$ point in Figure 1,
can be expected to arise in
a study of the relevant Clifford algebras.

\sect{11. Acknowledgments}
I thank Alan Kosteleck\'y for discussion.
Partial support for this work was provided by
the United States Department of Energy
(grant no.\ DE-FG02-91ER40661)
and by a Northern Michigan University
faculty research grant.

\sect{References}
\def\apb #1 #2 #3 {App. Phys. B {\bf #1}, #3, (19#2)} %
\def\ajp #1 #2 #3 {Am.\ J.\ Phys.\ {\bf #1}, #3 (19#2)}
\def\ant #1 #2 #3 {At. Dat. Nucl. Dat. Tables {\bf #1}, #3 (19#2)}
\def\ap #1 #2 #3 {Ann.\ Physics\ {\bf #1}, #3 (19#2)} %
\def\ijqc #1 #2 #3 {Internat.\ J.\ Quantum\ Chem.\
  {\bf #1}, #3 (19#2)} %
\def\jmp #1 #2 #3 {J.\ Math.\ Phys.\ {\bf #1}, #3 (19#2)}
\def\jms #1 #2 #3 {J.\ Mol.\ Spectr.\ {\bf #1}, #3 (19#2)} %
\def\jpa #1 #2 #3 {J.\ Phys.\ A {\bf #1}, #3 (19#2)}  %
\def\jpsj #1 #2 #3 {J.\ Phys.\ Soc.\ Japan {\bf #1}, #3 (19#2)}
\def\lnc #1 #2 #3 {Lett.\ Nuov.\ Cim. {\bf #1}, #3 (19#2)}
\def\mj #1 #2 #3 {Math. Japon. {\bf #1}, #3 (19#2)} %
\def\mpl #1 #2 #3 {Mod.\ Phys.\ Lett.\ A {\bf #1}, #3 (19#2)}
\def\nat #1 #2 #3 {Nature {\bf #1}, #3 (19#2)}
\def\nc #1 #2 #3 {Nuov.\ Cim.\ A{\bf #1}, #3 (19#2)}
\def\ncb #1 #2 #3 {Nuov.\ Cim.\ B{\bf #1}, #3 (19#2)}
\def\nima #1 #2 #3 {Nucl.\ Instr.\ Meth.\ Phys.\ Res.\
     A{\bf #1}, #3 (19#2)}
\def\nim #1 #2 #3 {Nucl.\ Instr.\ Meth.\ B{\bf #1}, #3 (19#2)}
\def\npb #1 #2 #3 {Nucl.\ Phys.\ B{\bf #1}, #3 (19#2)}
\def\pha #1 #2 #3 {Physica \ {\bf #1}, #3 (19#2)}
\def\pjm #1 #2 #3 {Pacific J. Math. \ {\bf #1}, #3 (19#2)} %
\def\pla #1 #2 #3 {Phys.\ Lett.\ A {\bf #1}, #3 (19#2)}
\def\plb #1 #2 #3 {Phys.\ Lett.\ B {\bf #1}, #3 (19#2)}
\def\prep #1 #2 #3 {Phys.\ Rep. {\bf #1}, #3 (19#2)}  %
\def\pra #1 #2 #3 {Phys.\ Rev.\ A {\bf #1}, #3 (19#2)}  %
\def\prd #1 #2 #3 {Phys.\ Rev.\ D {\bf #1}, #3 (19#2)}
\def\prdnew #1 #2 #3 {Phys.\ Rev.\ D {\bf #1}, #3 (20#2)}
\def\prl #1 #2 #3 {Phys.\ Rev.\ Lett.\ {\bf #1}, #3 (19#2)}
\def\prs #1 #2 #3 {Proc.\ Roy.\ Soc.\ (Lon.) A {\bf #1}, #3 (19#2)}
\def\ptp #1 #2 #3 {Prog.\ Theor.\ Phys.\ {\bf #1}, #3 (19#2)}
\def\rmp #1 #2 #3 {Rev.\ Mod.\ Phys.\ {\bf #1}, #3 (19#2)} %
\def\ibid #1 #2 #3 {\it ibid., \rm {\bf #1}, #3 (19#2)}  %
\def\baps #1 #2 #3 {Bull.\ Am.\ Phys.\ Soc.\ {\bf #1}, #3 (19#2)} %
\def\pp #1 #2 #3 {Phys.\ Plasmas\ {\bf #1}, #3 (19#2)} %
\def\ajm #1 #2 #3 {Am.\ J.\ Math. {\bf #1}, #3 (18#2)} %
\def\sciam #1 #2 #3 {Sci.\ Am. {\bf #1}, #3 (19#2)} %

%-------------------------------------------------------

\newpage
\begin{figure} % Fig 1.
\begin{center}
\epsfig{file=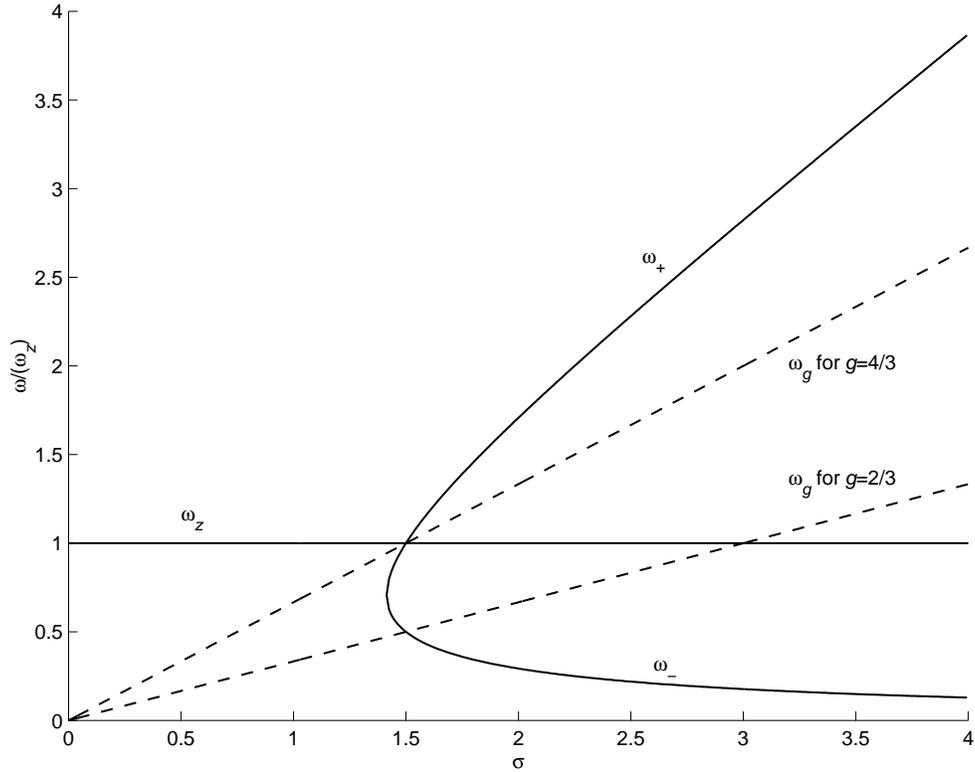, width=13cm}
\end{center}
\caption{
The four Penning-trap frequencies
$\om_+$, $\om_-$, $\om_g$, and $\om_z$
as functions of the parameter
$\si = \om_c / \om_z$.
The dashed lines show $\om_g$
for $g=4/3$ and $g=2/3$.
For $g=4/3$,
and there are three equal frequencies
$\om_+ = \om_z = \om_g = 2 \om_-$
at the supersymmetric point
$\si=3/2$.
For $g=3/2$ there are two pairs
of distinct equal frequencies at the
supersymmetric point.
The frequencies $\om_+$ and $\om_-$ have infinite slopes
where they meet at $\si=2^{1/2}$.
}
\end{figure}

\begin{figure} % Fig 2.
\begin{center}
\epsfig{file=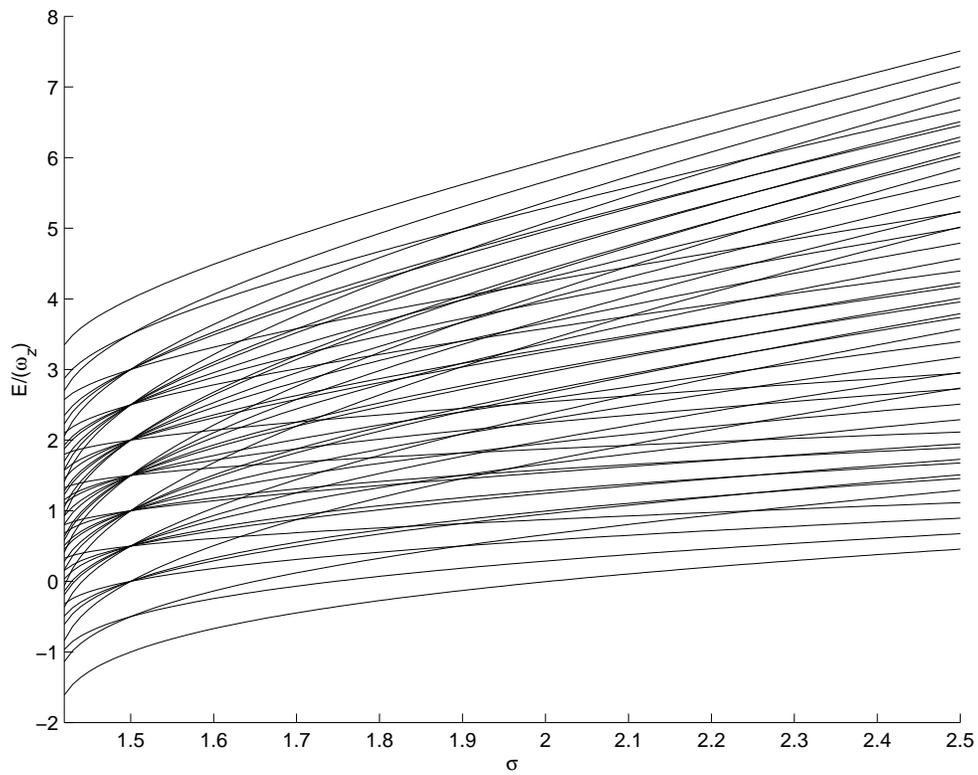, width=13cm}
\end{center}
\caption{
Penning-trap energies
as a function of
$\si$
for various states,
with $g=2/3$.
There are conspicuous degeneracies of the levels
at the supersymmetric point $\si=3/2$,
arising from
the superalgebra structure
discussed in section 6.
Another degeneracy occurs at $\si=2.25$.
In this plot, $\hb=1$.}
\end{figure}

\begin{figure}  % Fig 3
\begin{center}
\epsfig{file=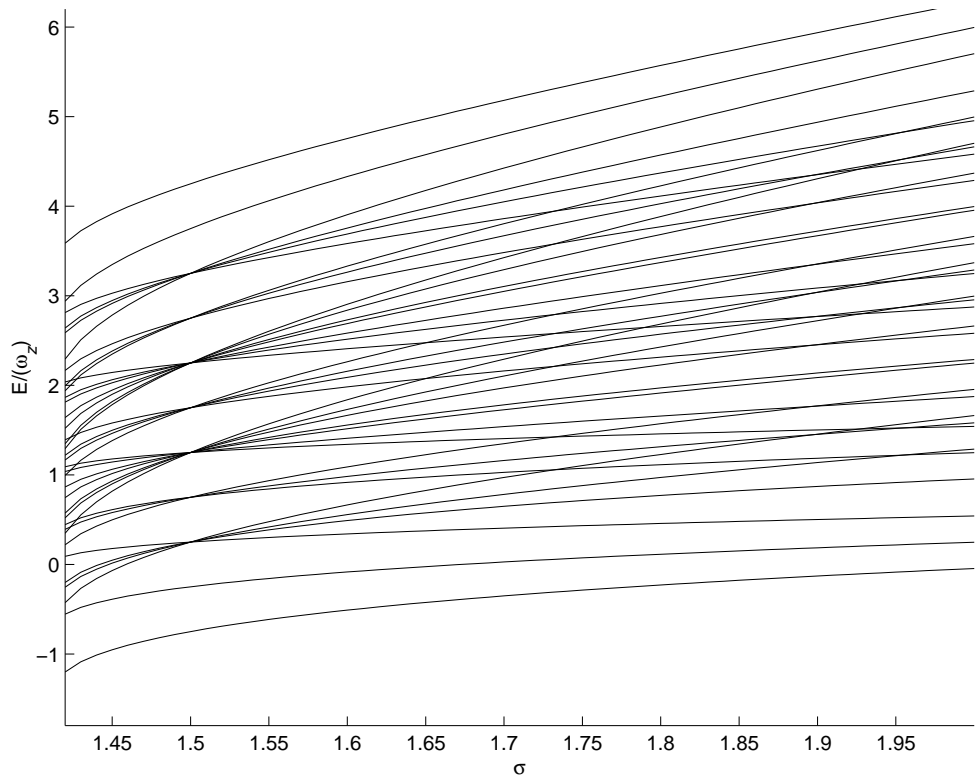, width=13cm}
\end{center}
\caption{Penning-trap
energies as a function of
$\si$
for various states,
with $g=4/3$.
The evenly spaced crossings
at the supersymmetric point $\si=3/2$
are discussed in section 7.
For this plot, $\hb=1$.}
\end{figure}

% =========================== start table ==================
\begin{figure}
\large
\begin{displaymath}
\begin{array}{||c||c|c|c|c||l||c||}
\hline\hline
g & \om_+ & \om_- & \om_z & \om_g &
\multicolumn{1}{c||}{\mbox{structure}}&\mbox{section} \\[1mm]
\hline\hline

 \fr 2 9     &  2 & \fr 1 4 & 1  & \fr 1 4 &
\alg u 1 \times \alg u 1 \times \alg u 1 \oslash \alg{su}{1|1}
& 5 \\[1mm]
\hline

 \fr 2 3     &  1 & \fr 1 2 & 1  & \fr 1 2 &
\alg u 1 \times\alg{so}{3}\times\alg u 1 \oslash \alg{su}{1|1}
& 6 \\[1mm]
\hline

\fr 4 3  &  1 & \fr 1 2 &  1& 1 &
\alg u 1 \times\alg u 1 \times\alg{su}{2|1} & 7 \\[1mm]
\hline

  &  1 &  1  &  1 & 1 &
\alg u 1 \times \alg{su}{2,1|1} & 8 \\[1mm]
\hline

\hline
\end{array}
\end{displaymath}
\normalsize
% ============================ end table ===================
%============================ caption ======================
\vspace*{1cm}

\noindent
{Table 1: }
Penning trap superalgebras
for the supersymmetric configuration $\si =3/2$.
The particle $g$ factor is given in the first column,
and the four frequencies in units of $\om_z$
are given in the next four columns.
The algebraic structures found
and the sections where they are discussed
are given in the final two columns.
The symbol $\oslash$ is defined in section 5.
The bottom row represents the hypothetical
case with four equal frequencies.
\end {figure}
%============================== end caption ==========

\end{document}